\documentclass[preprint,
preprintnumbers,amsmath,amssymb,superscriptaddress]{revtex4-2}

\usepackage{fullpage,amsthm,amsmath,amsfonts,bm,times,color,bbm}
\usepackage{graphicx}
\usepackage[english]{babel}
\usepackage{mathtools,color}

\usepackage{array}
\usepackage{float}
\usepackage{subfigure}
\usepackage{wrapfig}
\usepackage{hyperref}
\usepackage{romannum}
\usepackage{amsmath}
\usepackage{amssymb}
\usepackage{multirow}
\usepackage{booktabs}
\usepackage{makecell}
\usepackage[flushleft]{threeparttable}

\usepackage{soul} 

\makeatother

\makeatletter

\begin{document}
	
	
	\pagenumbering{arabic}
	
	\title{Global Patterns of Extreme Temperature Teleconnections Using Climate Network Analysis}%

	\author{Yuhao Feng}
	\affiliation{School of Science, Beijing University of Posts and Telecommunications, 100876 Beijing, China}
	\author{Jun Meng}
	\email{mengjun@mails.iap.ac.cn}
	\affiliation{State Key Laboratory of Earth System Numerical Modeling and Application, Institute of Atmospheric Physics, Chinese Academy of Sciences, Beijing, 100029, China}	
	\author{Jingfang Fan}%
	\email{jingfang@bnu.edu.cn}
	\affiliation{School of Systems Science/Institute of Nonequilibrium Systems, Beijing Normal University,  Beijing 100875, China}
	\affiliation{Potsdam Institute for Climate Impact Research, Potsdam 14412, Germany}
	
	\begin{abstract}
         Extreme weather events, rare yet profoundly impactful, are often accompanied by severe conditions. Increasing global temperatures are poised to exacerbate these events, resulting in greater human casualties, economic losses, and ecological destruction. Complex global climate interactions, known as teleconnections, can lead to widespread repercussions triggered by localized extreme weather. Understanding these teleconnection patterns is crucial for weather forecasting, enhancing safety, and advancing climate science. Here, we employ climate network analysis to uncover teleconnection patterns associated with extreme temperature fluctuations, including both extreme warming and cooling events occurring on a daily basis. Our study results demonstrate that the distances of significant teleconnections initially conform to a power-law decay, signifying a decline in connectivity with distance. However, this power-law decay tendency breaks beyond a certain threshold distance, suggesting the existence of long-distance connections. Additionally, we uncover a greater prevalence of long-distance connectivity among extreme cooling events compared to extreme warming events. The global pattern of teleconnections is, in part, driven by the mechanism of Rossby waves, which serve as a rapid conduit for inducing correlated fluctuations in both pressure and temperature.  These results enhance our understanding of the multiscale nature of climate teleconnections and hold significant implications for improving weather forecasting and assessing climate risks in a warming world.
	\end{abstract}
	\date{\today}
	\maketitle
\section{Introduction}
    Extreme weather events, invariably accompanied by severe weather conditions, are extraordinary and infrequent occurrences characterized by their low likelihood of happening but with profound and devastating consequences for human lives and society. Current observational evidence and climate model projections consistently indicate that rising global temperatures will intensify the frequency and severity of extreme weather events across the globe~\cite{tollefson2021ipcc,morit2022extreme,kumar2021climate,abbass2022review}, leading to a heightened magnitude of human casualties, extensive damage, increased economic burdens, and the destruction of ecosystems. Complex interactions and teleconnections within the global climate system pose potential threats to the world~\cite{lau20122010,boers2019complex}. Weather conditions, especially extreme events in one location, can have far-reaching effects, triggering a chain reaction of consequences on a global scale~\cite{walker1924correlations}. Hence, unraveling the teleconnection patterns of extreme weather events worldwide holds significant importance for weather forecasting, enhancing human safety, and gaining a deeper understanding of climate mechanisms~\cite{bridgman2014global,liu2007atmospheric}, particularly in the context of a warming global climate~\cite{tu_eigen_2025}. 
     
    The Earth's climate system consists of highly interconnected, nonlinear subsystems governed by intricate interactions and feedback mechanisms that span diverse spatial and temporal scales~\cite{fan2021statistical}. Employing a complex network approach to model the climate system offers valuable insights into its dynamic and interconnected behavior. Climate Network (CN) analysis, grounded in complex network theory, represents the climate system as a network in which nodes correspond to geographical locations and links encode statistical relationships—such as correlations or synchronizations—between climate variables at different regions~\cite{bridgman2014global,liu2007atmospheric}. Over the past decade, CNs have proven to be a versatile and powerful framework for analyzing climatic time series. For instance, Pearson correlation-based networks have yielded significant results in characterizing the El Niño–Southern Oscillation~\cite{meng2017percolation,meng2018forecasting,fan2017network}. Similarly, event synchronization networks have been successfully applied to investigate the spatiotemporal structure of extreme rainfall events~\cite{boers2019complex,boers2014prediction,boers2015complex}. Furthermore, mutual information-based networks have effectively captured the westward propagation of sea surface temperature (SST) anomalies associated with the Atlantic Multidecadal Oscillation, as demonstrated by Feng et al.~\cite{feng2014north}.

    In this study, we employ the event synchronization CN framework~\cite{boers2019complex} to investigate the spatial distribution of interactions among extreme weather events. We specifically concentrated on abrupt temperature changes and established criteria for extreme warming (positive) and cooling (negative) events by examining daily temperature variations compared to the previous day. These events are of particular interest due to their disruptive impacts on weather forecast accuracy and daily human activities, posing considerable risks to human safety and well-being. We find that extreme weather events occur less frequently over continents than over oceans, primarily due to differences in thermal dynamics. On a global scale, extreme events exhibit a relatively uniform distribution along longitudes but display distinct patterns along latitudes, reflecting the influential role of the jet stream in modulating weather phenomena such as storms. Within the framework of CN analysis, we identify both source regions that can significantly influence weather conditions elsewhere and sink regions that are particularly vulnerable to such external disruptions. Notably, our analysis reveals robust teleconnections between spatially distant extreme events, suggesting the potential for cascading failures within the climate system—potentially triggered through synchronization mechanisms. These findings emphasize the critical need to understand the complex interdependencies of Earth’s climate system to enhance the accuracy of climate modeling and forecasting.
	
\section{Results}
	
	\begin{figure}[]
		\centering
		\includegraphics[width=0.9\linewidth]{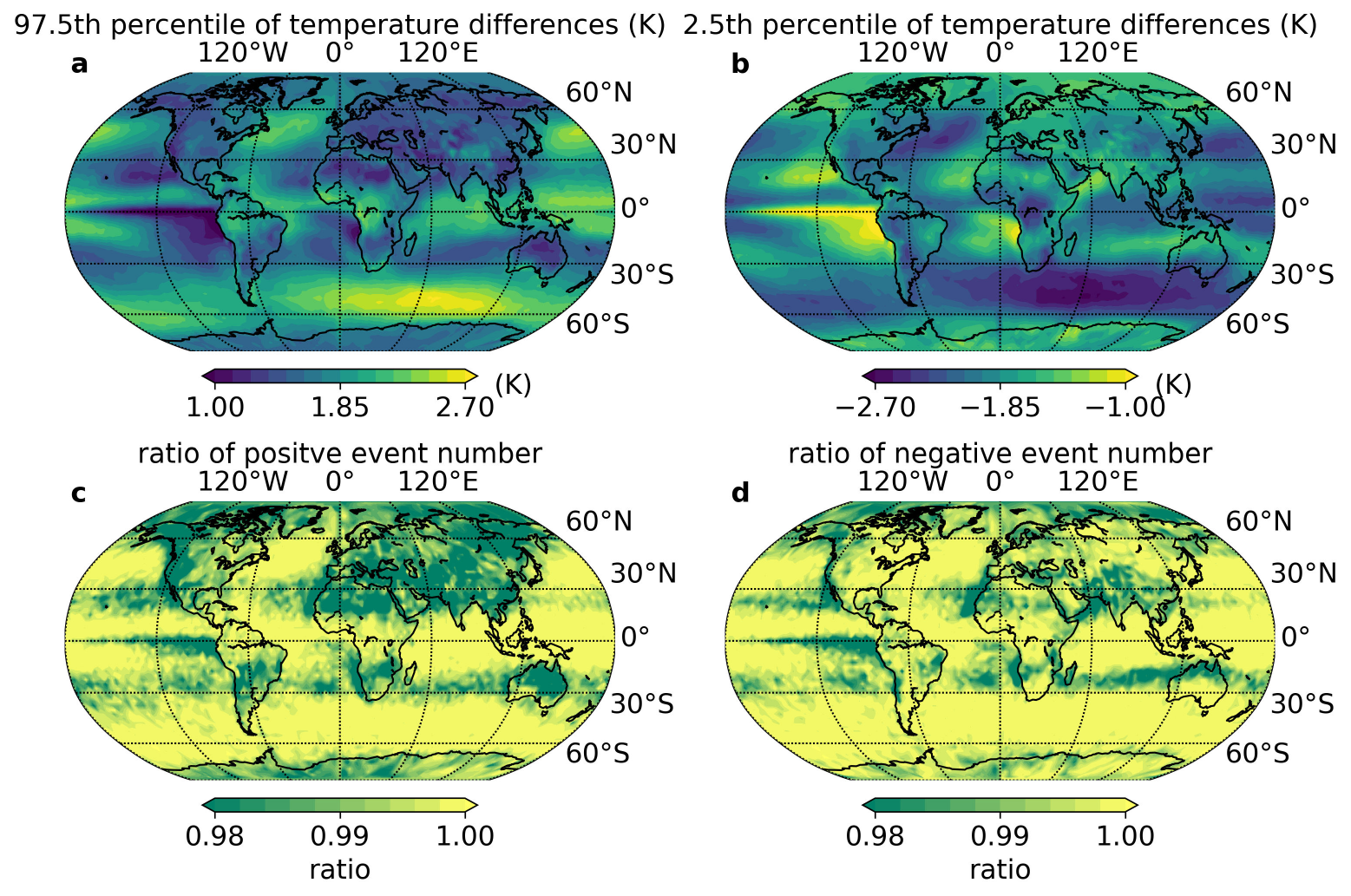}
		\caption{ { \bf Spatial patterns based on day-to-day air temperature differences by the threshold of $97.5\%$(pst) and $2.5\%$(ngt) changes.}
			{\bf a}, 
			Values of the 97.5th percentile of day-to-day air temperature differences from 1979 to 2021, computed individually for each of the 6570 timeseries obtained from the 2.5° × 2.5° grid cells of the ERA5 dataset. 
			{\bf b}, The same as a, but under the 2.5th percentile of day-to-day air temperature differences. As the difference approaches zero, it indicates a higher level of temperature stability within the respective region.
			{\bf c}, Total number of events above the threshold of the 97.5th percentile shown in {\bf a}. Air temperature differences above the threshold on consecutive days are treated as a single event. 
			{\bf d}, The same as c, but under the 2.5th percentile of day-to-day air temperature differences.
		}
		\label{fig1}%
	\end{figure}

    We utilize daily air temperature data at the 1000\,hPa pressure level from ERA5 reanalysis~\cite{hersbach2020era5}, with a spatial resolution of \(2.5^\circ \times 2.5^\circ\), encompassing 10,512 grid points globally. Due to the Earth's oblate spheroid shape, the actual ground distance represented by a given longitudinal interval decreases toward higher latitudes. To achieve a more uniform spatial sampling, we follow the method described in Section.~\ref{Universal}, selecting 6,570 nodes from the original set of 10,512 to ensure approximately equal-area coverage across the globe~\cite{meng2017percolation}. For each node (i.e., latitude–longitude grid point), we analyze daily temperature records from January 1, 1979, to December 31, 2021, after removing the mean seasonal cycle and normalizing by the corresponding seasonal standard deviation.
    
    To identify extreme events, we define node-specific thresholds using the 97.5th percentile for positive (extremely warm) events (Fig.~\ref{fig1}a) and the 2.5th percentile for negative (extremely cold) events (Fig.~\ref{fig1}b). Sequences of consecutive days exceeding these upper or lower thresholds are treated as as a single extreme event. A comprehensive description of the detection criteria is provided in Section.~\ref{datasection}. As shown in Fig.~\ref{fig1}a and~\ref{fig1}b, the upper and lower temperature thresholds are less pronounced near the equator than at higher latitudes, especially over oceans. This indicates greater temperature stability at lower latitudes, where equatorial temperatures remain nearly constant year-round, varying by only a few degrees~\cite{harvey2014equator}. Additionally, Fig.\ref{fig1}c and~\ref{fig1}d reveal that regions with fewer detected extreme events are predominantly continental. However, this does not imply that extreme weather is less frequent over land than over oceans. Rather, the reduced event count on land is primarily due to event clustering~\cite{perkins2017changes}, where sequences of consecutive days with abrupt temperature changes are aggregated as a single event. This contrast arises from the distinct thermal dynamics of land and ocean surfaces. Oceans, with their higher heat capacity, absorb and retain heat over extended periods, resulting in more temporally dispersed extreme events. In contrast, land surfaces, which heat and cool more rapidly due to lower heat capacity, tend to experience more frequent and tightly grouped extreme events. Furthermore, the observed lower frequency of extreme events in mid-latitudes suggests that the jet stream plays a critical role in organizing storm systems, thereby promoting the clustering of temperature anomalies in these regions~\cite{rousi2022accelerated}.
    
    We quantify connections among extreme events using the event synchronization CN method (see Section.~\ref{synchronization} and ~\ref{Universal}). Network links are established between nodes when synchronization values are significant ($P < 0.005$; see Section.~\ref{Significance tests}). Each node’s degree, defined as the number of significant links, reflects its interaction with the global climate system—higher degrees indicate stronger connectivity, while lower degrees suggest isolation. Using this approach, we constructed three CNs: the positive-to-positive (PP) network, based on synchronized positive events; the negative-to-negative (NN) network, based on synchronized negative events; and the positive-to-negative (PN or NP) network, based on synchronized positive (negative) and negative (positive) events.

\begin{figure}[h!]
		\centering
		\includegraphics[width=1.0\linewidth]{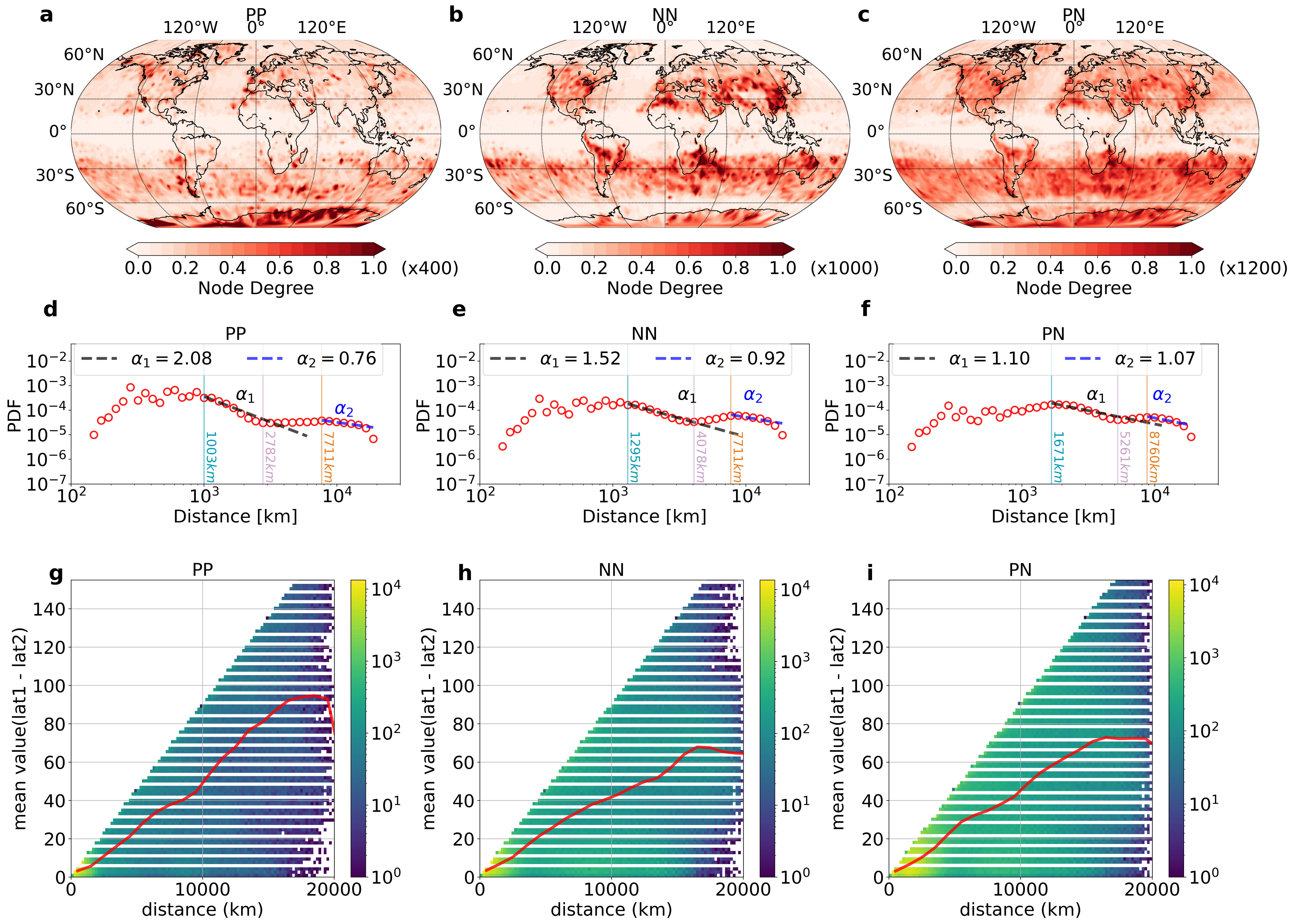}
		\caption{ { \bf Characteristics of climate networks based on extreme-event synchronizations.}
            \textbf{a, b, c}, Spatial distribution of node degrees for the positive-positive (PP), negative-negative (NN), and positive-negative (PN) climate networks, respectively. \textbf{d, e, f}, Probability density functions (PDFs) of link distances, defined as the great-circle distance between pairs of nodes connected by significant links. Power-law fits are shown as dashed black and blue lines over the relevant range. \textbf{g, h, i}, Latitude difference analysis of synchronized node pairs. Each panel shows the absolute difference in latitude (in degrees) between linked nodes. Red lines indicate the mean latitude difference within each bin.
            }
		\label{fig2}%
	\end{figure}

    Our findings, presented in Fig.~\ref{fig2}(a–c), reveal a striking pattern: although extreme events are most frequent near the equator (see Fig.~\ref{fig1}c, d), these equatorial nodes exhibit near-zero degrees—indicating limited connectivity with the broader climate network. In contrast, polar and mid-latitude regions—particularly continental areas such as the Amazon, Southeast China, Eastern Australia, and Northern Africa—demonstrate significantly higher node degrees, suggesting stronger interactions within the climate system. Notably, negative extreme events show greater overall connectivity: the NN and PN networks display markedly higher node degrees than the PP network (Fig.~\ref{fig2}b, c vs. Fig.~\ref{fig2}a). These observations raise two critical questions: (1) What are the spatial origins of the significant links associated with high-degree nodes, and how do they relate to regional influence? (2) What is the directionality of these links—do they signify a node's role as a source of disruption (influential) or as a recipient (vulnerable)?
    
    To address the first question, we analyze the Probability Density Function (PDF) of distances for significant links. Fig.~\ref{fig2}(d-f) reveals a multiscale teleconnection structure. At local spatial scales (1000–3000 km), all three network types (PP, NN, PN) show a power-law decay, reflecting a rapid drop in synchronization likelihood with distance. in link probability, indicating a rapid decline in synchronization likelihood with increasing distance. However, beyond approximately 4,000 km, the distributions deviate from this scaling behavior, showing enhanced tails or secondary peaks. These long-range teleconnections correspond to``dragon kings”~\cite{sornette2012dragon, sachs2012black}—extreme, statistically anomalous events that stand out from the expected short-range scaling law. This finding suggests that remote synchronization between extreme events arises from underlying processes that cannot be inferred from local dynamics alone, highlighting the presence of emergent global-scale organizing mechanisms in the climate system~\cite{wang_exploring_2023}.

    High-degree nodes are predominantly clustered in mid-latitude and polar regions (Fig.~\ref{fig2}a–e), aligning with jet stream pathways and the propagation of Rossby waves, which facilitate the long-range transport of atmospheric perturbations across thousands of kilometers. For instance, the NN network’s secondary peak at 6000-8000 km (Fig.~\ref{fig2}e) matches typical Rossby wave wavelengths, implicating them in long-distance cold air outbreaks and synoptic disturbances~\cite{boers2019complex,zhang2019significant}. This interplay likely drives synchronization across vast distances, such as linking polar cold outbreaks with downstream cooling.  Furthermore, Fig.~\ref{fig2}(g–i) reveals that long-distance synchronization often involves cross-latitude and even cross-hemispheric connectivity, which is consistent with the influence of Rossby waves and jet stream dynamics in facilitating energy transport across climatic zones. Overall, the multiscale patterns in Fig.~\ref{fig2} reflect dual mechanisms: local high-frequency coupling (short-range power-law decay) and remote wave-driven propagation (long-range Rossby wave effects).

	\begin{figure}[hbpt]
		\centering
		\includegraphics[width=1\linewidth]{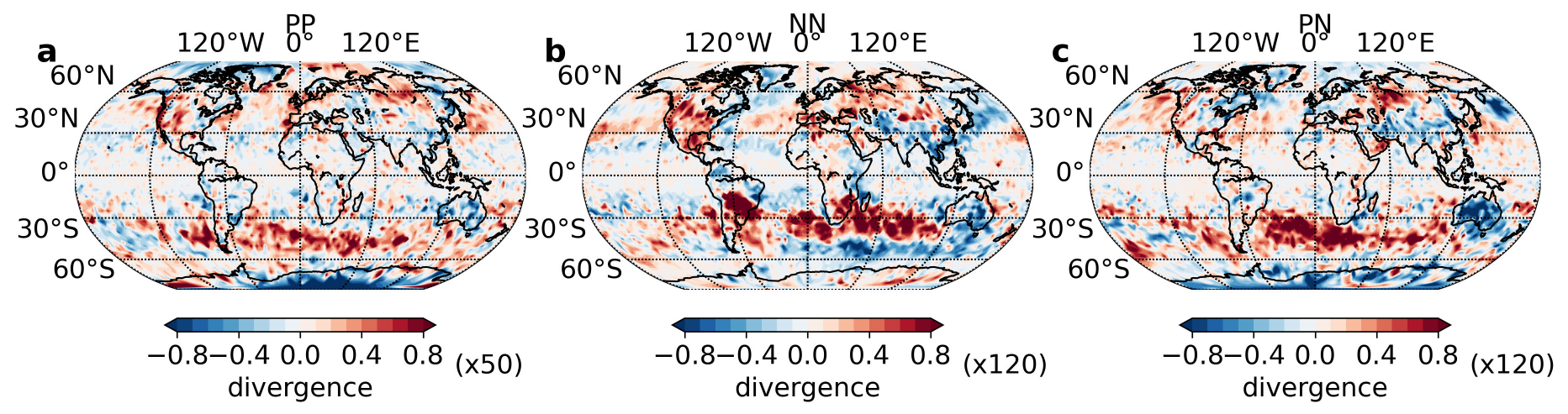}
		\caption{ { \bf Some characteristics about directed climate network of extreme-event synchronizations.}
			{\bf a, b, c} Spatial distribution of divergence for positive-to-positive (PP), negative-to-negative (NN), and positive-to-negative (PN) climate networks, respectively. Divergence, defined as the difference between out-degree and in-degree for each grid cell, is standardized by dividing by the maximum absolute divergence value to enable meaningful comparisons across networks. Color scales range from -0.8 to 0.8 (PP: ×50; NN: ×120; PN: ×120), with positive values (red) indicating a node’s influence and negative values (blue) indicating vulnerability.
		}
		\label{fig3}%
	\end{figure}
    
    To address the second question, we incorporate directional information into the original CNs by determining the direction of each link based on time lags of synchronized events between nodes (see Section.~\ref{high synchronization}). We define a node’s out-degree as the number of outgoing links and its in-degree as the number of incoming links (see Data and Methods~\ref{Climate networks}). The out-degree reflects the node’s influence on the global climate system, while the in-degree indicates its susceptibility to external impacts. We further introduce the concept of divergence, calculated as the out-degree minus the in-degree. A high positive divergence signifies a node’s strong influence on the climate system, whereas a high negative divergence indicates greater vulnerability. Network divergence enables the identification of influential or vulnerable regions within the system.

    Our analysis (Fig.~\ref{fig3}) reveals a pronounced concentration of high divergence in the southern mid-to-high latitudes (20°S–60°S), indicating that this region functions predominantly as a net source of extreme temperature events—actively propagating signals rather than passively receiving them. For the PP network (Fig.~\ref{fig3}a), when Southern Hemisphere westerlies shift northward or intensify, enhanced warm air advection strengthens heat exchange between midlatitude warm oceanic regions—especially near the northern boundary of the Southern Ocean—and the overlying atmosphere. This process, coupled with planetary-scale or regional circulation anomalies, can trigger extreme warming events and propagate ``warming signals" downstream~\cite{cai2012more, goyal2021historical}. In the NN network (Fig.~\ref{fig3} b), a negative phase of the Southern Annular Mode displaces westerlies northward, facilitating the advection of polar cold air and moisture. As polar air intrudes via cyclonic circulations, storm tracks, and low-pressure systems, cooling signals are efficiently transmitted to lower latitudes~\cite{meehl2007global,rintoul2018global}. For the PN network (Fig.~\ref{fig3}c), phase shifts in background flow—such as changes in westerly speed or intensity—modify the propagation characteristics of Rossby waves. These alterations induce abrupt transitions between warming and cooling events across different longitudes within the same latitude band, reflecting dynamic interactions between opposing extreme events~\cite{tsonis2008topology,donges2009complex}.

\begin{figure}[hbpt]
		\centering
        \includegraphics[width=0.7\linewidth]{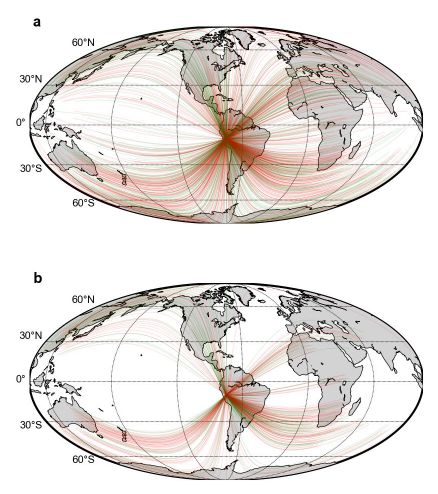}
		\caption{ { \bf Out-links and in-links of negative-to-negative (NN) climate networks  in the Amazon (red lines denote out-edges, green lines denote in-edges)}. 
        {\bf a.} Uncorrected out-links and in-links network;
        {\bf b.} Corrected out-links and in-links network.
		}
            \label{fig4}
	\end{figure}

    Moreover, Fig.\ref{fig3} demonstrates that the importance of high-divergence regions extends beyond their immediate influence: we observe a notable spatial overlap between regions exhibiting high divergence values (±0.6 to ±0.8) and key climate tipping elements, including the Antarctic Ice Sheet, North American and Eurasian boreal forests, and the Amazon rainforest~\cite{lenton2008tipping,lenton2019climate}. This spatial coincidence suggests that state transitions in these regions—particularly the onset of tipping points—may not only cause localized disruptions but also amplify and trigger extreme events that propagate globally through the synchronization network of extremes. Motivated by this insight and the consistently high divergence values observed in Fig.~\ref{fig2} (ranging from 0.6 to 1), we focus on the Amazon rainforest as a case study to further investigate the teleconnection structure associated with extreme cold events (see Fig.~\ref{fig4}).
        
    Before proceeding, we address a fundamental statistical challenge inherent in all data-driven correlation analyses, particularly when constructing networks based on statistical similarity. These methods are highly susceptible to multiple comparison bias. In our case, each time series was compared with 6,569 others, yielding over \(10^8\) pairwise comparisons. Consequently, the resulting network may contain spurious links that arise purely by chance, rather than reflecting genuine physical connections. This issue is particularly relevant for the spatial distribution of network connections linked to the Amazon rainforest (see Fig.~\ref{fig4}a). To mitigate this, we apply a spatial density estimation method (see Section.~\ref{Bias}) to identify statistically significant (\(P < 0.001\)) spatial link bundles. This procedure effectively accounts for multiple testing within the spatially embedded functional network. All links outside these statistically validated bundles are removed, allowing us to isolate physically meaningful couplings with greater confidence.

    After correction, the refined data reveal a robust pattern of stable link bundles between the Amazon rainforest and distant regions, including Europe, eastern Australia, tropical Africa, and the mid-latitudes of the Southern Hemisphere. As shown in Fig.~\ref{fig4}b, these connections, depicted as red (out-degree) and green (in-degree) edges during extreme cold events in the Amazon, capture reciprocal teleconnections across specific spatial and temporal scales. This distinctive network structure provides direct empirical evidence for examining how large-scale atmospheric waves mediate the propagation of regional extreme climate events.  Strikingly, this edge pattern aligns well with known atmospheric teleconnection pathways governed by Rossby waves. As dominant features of mid-to-high latitude atmospheric dynamics, Rossby waves exhibit characteristic wavelength ranges that closely match the spatial distances of these links. This suggests that during extreme cold events in the Amazon, cold-air anomalies may be transmitted to Europe, eastern Australia, and tropical Africa via Rossby wave trains. Simultaneously, the Amazon also receives feedback from these regions, forming a complex bidirectional interaction. Thus, the Amazon functions as both a \textit{driver} and \textit{responder} of influence within the global extreme event network, underscoring its pivotal and dynamic role in the Earth’s climate system~\cite{nobre2016land}.
        	
	\section{Discussions}
  
    In this study, we have developed a comprehensive CN approach to unravel the teleconnection patterns of extreme temperature events. By leveraging event synchronization techniques, our work captures both local interactions and remote, long-range connections that are predominantly mediated by atmospheric wave dynamics.
    
    One of the main findings is the power-law decay in the probability density of link distances at scales up to approximately 4000 km. This decay indicates that the likelihood of two regions experiencing synchronized extreme events decreases rapidly with distance. However, the observed deviation beyond this threshold points to the existence of a secondary mechanism—likely associated with the propagation of Rossby waves—which facilitates robust teleconnections over intercontinental distances. The emergence of “dragon king” type anomalies in the link distance distribution underscores that remote regions can become unexpectedly connected, a phenomenon that has significant implications for our understanding of climate variability.
    
    Furthermore, our directed network analysis sheds light on the roles of individual nodes within the global climate system.  Regions exhibiting high positive divergence are identified as key drivers capable of influencing distant areas through rapid atmospheric processes.  In contrast, high positive divergence nodes (e.g., mid-latitude oceans) act as sources of extreme event propagation, while high negative divergence areas (e.g., polar regions) are vulnerable to external perturbations. The overlap between these high-divergence zones and climate tipping elements (e.g., Amazon rainforest) is a pivotal discovery, suggesting that disturbances in these “conduit regions” could trigger cascading failures via atmospheric pathways. For example, extreme cooling in the Amazon was linked to synchronized events in Europe and Australia, mediated by Rossby wave trains, highlighting the potential for remote tipping element interactions.

    Overall, our results emphasize the dual role of local thermodynamic processes and large-scale atmospheric dynamics in governing extreme temperature teleconnections. As global warming intensifies, the frequency and impact of these events are likely to increase, making it imperative to incorporate such network-based insights into future climate forecasting and risk management strategies. By bridging the gap between short-term variability and long-term climate change, our study provides a more nuanced understanding of the interconnected nature of extreme weather phenomena and highlights the urgent need to monitor and mitigate potential cascading effects in a warming world.
	
	\clearpage
	
	\section{Data and Methods}
	\subsection{Data} \label{datasection}

    The data utilized in this study were extracted from the ERA5 reanalysis datasets, available through the European Centre for Medium-Range Weather Forecasts (ECMWF) website~\cite{hersbach2020era5}. The data exhibits a spatial resolution of 2.5 degrees in both the zonal (latitude) and meridional (longitude) directions. This study predominantly employs the air temperature data at the 1000hPa pressure level at 0 hours (UTC). The 1000hPa pressure level air temperature represents the atmospheric temperature at a specific vertical elevation above the Earth's surface.
    
    We define day-to-day air temperature differences as the current day's air temperature minus the previous day's at each gird from 1979 to 2021. To ensure consistency in the calendar year's duration and simplify the analysis, we excluded leap days. For each of the 6,570 girds, we obtained a time series with a length of 15,694. Based on these, positive changes (extremely warm) are defined as those that are above the 97.5th percentile of daily air temperature variations. Similarly, values under the 2.5th percentile of day-to-day air temperature differences are considered negative changes (extremely cold).

	\subsection{Synchronization of extreme events.}\label{synchronization} 

    The ES algorithm operates as follows: Given a percentile threshold $p$, we denote the set of events above $p$ at grid cell $i$ as $\left\{{ }^{p} e_{i}^{\mu}\right\}{ }_{\mu=1, \ldots ., l_{i}}$, where $l_{i}$ represents the total number of events at grid cell $i$. Consecutive events occurring on the same day are counted as a single event. For each pair of grid cells $(i, j)$, ES calculates the count of pairs of uniquely associable events. To ensure uniqueness, we impose the condition that the absolute value of the temporal delay between any two synchronous events $\mu$ and $\nu$, denoted as $t_{i, j}^{\mu, \nu}:={ }^{p} e_{i}^{\mu}-^{p} e_{j}^{\nu}$, must be smaller than  $\tau_{i, j}^{\mu, \nu}:=\frac{\min \left\{t_{i, i}^{\mu, \mu-1}, t_{i, i}^{\mu+1, \mu}, t_{j, j}^{\nu, \nu-1}, t_{j, j}^{\nu+1, \nu}\right\}}{2}$. We set a maximum temporal delay of $\tau_{\max }=30$ days between events at different locations. Thus, ES for the pair of locations $(i, j)$ is defined as the number of event pairs $(\mu, \nu)$ that satisfy these conditions:
		
	\begin{equation}
		\mathrm{ES}_{i j}:=\left\|\left\{(\mu, \nu):\left|t_{i, j}^{\mu, \nu}\right|<\tau_{i, j}^{\mu, \nu} \wedge\left|t_{i, j}^{\mu, \nu}\right| \leq \tau_{\max }\right\}\right\|
	\end{equation}	
    Here, \( |a| \) represents the absolute value of a scalar quantity \( a \), and \( \|M\| \) denotes the cardinality of a set \( M \). This surrogate model preserves the total number of extreme events in each time series while randomizing their timing, thereby enabling rigorous significance testing~\cite{boers2019complex}.

	\subsection{Directional synchronization of extreme events.}\label{high synchronization} 
	  
    One key advantage of the event synchronization (ES) measure is its ability to incorporate a dynamic delay \( \tau_{i, j}^{\mu, \nu} \) within the range \( \tau_{i, j}^{\mu, \nu} \), unlike the fixed delay used in traditional lead-lag correlation analyses. Additionally, an adaptation of this measure allows us to pinpoint the exact times when extreme event synchronization between two regions of interest is high, while maintaining the temporal sequence. For two sets of time series, \( A \) and \( B \), associated with two different regions, we define:
    
	\begin{equation}
		\mathrm{ES}_{A \rightarrow B}^{\mu}:=\left\|\left\{(i, j) \in A \times B:-\tau_{i, j}^{\mu, \nu}<t_{i, j}^{\mu, \nu} \leq 0 \wedge\left|t_{i, j}^{\mu, \nu}\right| \leq \tau_{\max }\right\}\right\|
	\end{equation}
	and
	\begin{equation}
		\mathrm{ES}_{B \rightarrow A}^{\nu}:=\left\|\left\{(i, j) \in A \times B: 0 \leq t_{i, j}^{\mu, \nu}<\tau_{i, j}^{\mu, \nu} \wedge\left|t_{i, j}^{\mu, \nu}\right| \leq \tau_{\max }\right\}\right\|
	\end{equation}
	Here, \( A \times B \) represents the Cartesian product of sets \( A \) and \( B \), which includes all possible pairs \( (i, j) \) such that \( i \in A \) and \( j \in B \).
	
	Therefore, \( ES_{A \rightarrow B}^{\mu}(ES_{B \rightarrow A}^{\nu}) \) can be interpreted as a time series that indicates, for each time step, the number of events in region \( A \) (or \( B \)) that have a subsequent, uniquely corresponding event in region \( B \) (or \( A \)).
	
	\subsection{Climate networks}\label{Climate networks}

    We utilized the ES method to establish CNs encompassing both directed and undirected CN elements. For each pair of nodes, a network link is established if the corresponding synchronization value is statistically significant at \( P \leq 0.005 \).
 
    \subsubsection{Degree}\label{Degree} 
    
    In the realm of climate networks, a node's degree serves as a pivotal metric, representing the number of connections or links it has with other nodes. This metric is fundamental as it not only quantifies a node's connectivity but also provides profound insights into its significance and role within the overall network structure. In the context of our climate network, we specifically interpret a node's degree as the number of other grid cells that exhibit event synchrony with a particular grid cell.
    
    \subsubsection{In-degree \& Out-degree}
    In a directed climate network, we adopt two distinct yet complementary metrics to comprehensively describe the degree of node \(i\):
    - In-degree (\(k_i^{\text{in}}\)): The count of incoming edges directed towards node \(i\), signifying the number of grid cells that synchronize events with node \(i\).
    - Out-degree (\(k_i^{\text{out}}\)): The count of outgoing edges originating from node \(i\), representing the number of grid cells that synchronize events from node \(i\).
    
    The total degree \(k_i\) of node \(i\) is defined as:
    \[
    k_i = k_i^{\text{in}} + k_i^{\text{out}}
    \]
    This combined measure captures the overall connectivity strength of the node, regardless of directional bias.
    
    \subsubsection{Divergence}
    To characterize directional imbalances in node influence, we introduce the \textbf{divergence} metric:
    \[
    D_i = k_i^{\text{out}} - k_i^{\text{in}}
    \]
    - \(D_i > 0\): Indicates net outgoing influence (source node)
    - \(D_i < 0\): Indicates net incoming influence (sink node)
    - \(D_i = 0\): Balanced influence (neutral node)
    
    Divergence quantifies the asymmetric role of nodes in propagating extreme event synchronizations, highlighting regions with disproportionate influence over global teleconnections.
    
    \subsubsection{Link Distance Distribution}
    The link distance distribution is defined as the probability density function (PDF) of Euclidean distances between all pairs of connected nodes in the network. 
	\subsection{Significance tests}\label{Significance tests}
	
	The significance of event synchronization and network construction is determined as follows. To assess the statistical significance of each observed $ES_{ij}$ value, a null-model distribution is generated through numerical computations. This involves calculating ES for 2,000 pairs of surrogate event series, where the event numbers $l_i$ and $l_j$ are randomly and uniformly distributed. For each pair $(l_i, l_j)$, the 99.5th percentile of the corresponding distribution is identified as the significance threshold. Based on this null model, a network link is established between locations $i$ and $j$ if the value of $ES_{ij}$ surpasses the significance threshold, indicating its significance at a level of 0.005. It is important to note that by incorporating network links based on the statistical significance of $ES_{ij}$, potential biases arising from different event rates are eliminated.

    \subsection{finite-size analysis of the climate network}\label{Universal}

    It's worth noting that these nodes do not uniformly cover the entire globe due to the Earth's ellipsoidal shape. To address this non-uniformity, we employ a procedure developed by us~\cite{meng2017percolation}. We define the resolution (in degrees latitude) at the Equator as \(r_{0}\) and compute the number of nodes \(n_{0} = 360 / r_{0}\). Subsequently, the number of nodes at latitude \(m r_{0}\) is given by \(n_{m} = n_{0} \cos (m r_{0})\), where \(m \in [-90 / r_{0}, 90 / r_{0}]\). The total number of nodes is then calculated as \(N^{\prime} = \sum_{m=0}^{m=90 / r_{0}} (2 n_{m} - n_{0})\). Here, we set \(r_{0}\) to \(7.5^{\circ}\), resulting in \(N^{\prime} = 726\).

	\subsection{Bias Correction Method}\label{Bias}

    As elaborated in the main text, outcomes from data - driven interdependency analyses--especially those related to link configurations within functional networks--are typically susceptible to biases induced by multiple comparisons. Herein, we introduce an alternative method to rectify such biases, capitalizing on the spatially embedded nature of the system under scrutiny~\cite{boers2019complex}. The fundamental premise is that links governed by physical mechanisms should manifest spatially coherent patterns, distinguishing them from links originating from random coincidences.

    This method utilizes a null model to identify regions forming significant link bundles. To construct a statistical null model for regional link density, we first randomly distribute an equivalent number of links across all grid cells. Subsequently, Gaussian Kernel Density Estimation (KDE) is applied to quantify spatial link density, with the bandwidth determined by Scott’s rule and the Haversine metric employed to handle spherical geographic coordinates. By repeating this procedure 1,000 times, a null model distribution is generated for each grid cell, which serves as the baseline for identifying significant link bundles. Specifically, any grid cell where the observed regional link density exceeds the $99.9$th percentile of the null model distribution (i.e., \(P < 0.001\)) is classified as part of a significant link bundle.
    
	
	\section*{code availability}
	The Python codes used for the analysis are available on GitHub (\url{https://github.com/AaronFeng29/tas_ere_tele}).
	
	{\section*{Author Contributions}
		J.F and J.M designed the research. Y.F, J.M and J.F performed the analysis and prepared the manuscript, Y.F, J. M and J.F discussed results, and contributed to writing the manuscript. J.M and J.F led the writing of the manuscript. }
	
{\section*{Acknowledgements}}
This work was supported by the National Natural Science Foundation of China (Grant No. 42450183, 12275020, 12135003, 12205025, 42461144209), the Ministry of Science and Technology of China (2023YFE0109000). J.F. is supported by the Fundamental Research Funds for the Central Universities.
	
	\section*{Competing interests}
	The authors declare no competing interests.
	
	\clearpage
	\bibliographystyle{naturemag}
	\bibliography{main20250325}

\begin{thebibliography}{10}
\expandafter\ifx\csname url\endcsname\relax
  \def\url#1{\texttt{#1}}\fi
\expandafter\ifx\csname urlprefix\endcsname\relax\def\urlprefix{URL }\fi
\providecommand{\bibinfo}[2]{#2}
\providecommand{\eprint}[2][]{\url{#2}}

\bibitem{tollefson2021ipcc}
\bibinfo{author}{Tollefson, J.} \emph{et~al.}
\newblock \bibinfo{title}{Ipcc climate report: Earth is warmer than it’s been
  in 125,000 years}.
\newblock \emph{\bibinfo{journal}{Nature}} \textbf{\bibinfo{volume}{596}},
  \bibinfo{pages}{171--172} (\bibinfo{year}{2021}).

\bibitem{morit2022extreme}
\bibinfo{author}{MORIT, A.}
\newblock \bibinfo{title}{Extreme heatwaves: Surprising lessons from the record
  warmth}.
\newblock \emph{\bibinfo{journal}{Nature}} \textbf{\bibinfo{volume}{608}}
  (\bibinfo{year}{2022}).

\bibitem{kumar2021climate}
\bibinfo{author}{Kumar, P.}
\newblock \bibinfo{title}{Climate change and cities: challenges ahead}
  (\bibinfo{year}{2021}).

\bibitem{abbass2022review}
\bibinfo{author}{Abbass, K.} \emph{et~al.}
\newblock \bibinfo{title}{A review of the global climate change impacts,
  adaptation, and sustainable mitigation measures}.
\newblock \emph{\bibinfo{journal}{Environmental Science and Pollution
  Research}} \textbf{\bibinfo{volume}{29}}, \bibinfo{pages}{42539--42559}
  (\bibinfo{year}{2022}).

\bibitem{lau20122010}
\bibinfo{author}{Lau, W.~K.} \& \bibinfo{author}{Kim, K.-M.}
\newblock \bibinfo{title}{The 2010 pakistan flood and russian heat wave:
  Teleconnection of hydrometeorological extremes}.
\newblock \emph{\bibinfo{journal}{Journal of Hydrometeorology}}
  \textbf{\bibinfo{volume}{13}}, \bibinfo{pages}{392--403}
  (\bibinfo{year}{2012}).

\bibitem{boers2019complex}
\bibinfo{author}{Boers, N.} \emph{et~al.}
\newblock \bibinfo{title}{Complex networks reveal global pattern of
  extreme-rainfall teleconnections}.
\newblock \emph{\bibinfo{journal}{Nature}} \textbf{\bibinfo{volume}{566}},
  \bibinfo{pages}{373--377} (\bibinfo{year}{2019}).

\bibitem{walker1924correlations}
\bibinfo{author}{Walker, G.~T.}
\newblock \bibinfo{title}{Correlations in seasonal variations of weather. viii,
  a further study of world weather}.
\newblock \emph{\bibinfo{journal}{Men. Indian Meteor. Dept.}}
  \textbf{\bibinfo{volume}{24}}, \bibinfo{pages}{275--332}
  (\bibinfo{year}{1924}).

\bibitem{bridgman2014global}
\bibinfo{author}{Bridgman, H.~A.} \& \bibinfo{author}{Oliver, J.~E.}
\newblock \emph{\bibinfo{title}{The global climate system: patterns, processes,
  and teleconnections}} (\bibinfo{publisher}{Cambridge University Press},
  \bibinfo{year}{2014}).

\bibitem{liu2007atmospheric}
\bibinfo{author}{Liu, Z.} \& \bibinfo{author}{Alexander, M.}
\newblock \bibinfo{title}{Atmospheric bridge, oceanic tunnel, and global
  climatic teleconnections}.
\newblock \emph{\bibinfo{journal}{Reviews of Geophysics}}
  \textbf{\bibinfo{volume}{45}} (\bibinfo{year}{2007}).

\bibitem{tu_eigen_2025}
\bibinfo{author}{Tu, H.} \emph{et~al.}
\newblock \bibinfo{title}{Eigen microstate analysis unveils climate dynamics}.
\newblock \emph{\bibinfo{journal}{SCIENCE CHINA Physics, Mechanics \&
  Astronomy}} \textbf{\bibinfo{volume}{68}}, \bibinfo{pages}{240511}
  (\bibinfo{year}{2025}).
\newblock \bibinfo{note}{Publisher: Science China Press}.

\bibitem{fan2021statistical}
\bibinfo{author}{Fan, J.} \emph{et~al.}
\newblock \bibinfo{title}{Statistical physics approaches to the complex earth
  system}.
\newblock \emph{\bibinfo{journal}{Physics reports}}
  \textbf{\bibinfo{volume}{896}}, \bibinfo{pages}{1--84}
  (\bibinfo{year}{2021}).

\bibitem{meng2017percolation}
\bibinfo{author}{Meng, J.}, \bibinfo{author}{Fan, J.},
  \bibinfo{author}{Ashkenazy, Y.} \& \bibinfo{author}{Havlin, S.}
\newblock \bibinfo{title}{Percolation framework to describe el ni{\~n}o
  conditions}.
\newblock \emph{\bibinfo{journal}{Chaos: An Interdisciplinary Journal of
  Nonlinear Science}} \textbf{\bibinfo{volume}{27}} (\bibinfo{year}{2017}).

\bibitem{meng2018forecasting}
\bibinfo{author}{Meng, J.}, \bibinfo{author}{Fan, J.},
  \bibinfo{author}{Ashkenazy, Y.}, \bibinfo{author}{Bunde, A.} \&
  \bibinfo{author}{Havlin, S.}
\newblock \bibinfo{title}{Forecasting the magnitude and onset of el ni{\~n}o
  based on climate network}.
\newblock \emph{\bibinfo{journal}{New Journal of Physics}}
  \textbf{\bibinfo{volume}{20}}, \bibinfo{pages}{043036}
  (\bibinfo{year}{2018}).

\bibitem{fan2017network}
\bibinfo{author}{Fan, J.}, \bibinfo{author}{Meng, J.},
  \bibinfo{author}{Ashkenazy, Y.}, \bibinfo{author}{Havlin, S.} \&
  \bibinfo{author}{Schellnhuber, H.~J.}
\newblock \bibinfo{title}{Network analysis reveals strongly localized impacts
  of el ni{\~n}o}.
\newblock \emph{\bibinfo{journal}{Proceedings of the National Academy of
  Sciences}} \textbf{\bibinfo{volume}{114}}, \bibinfo{pages}{7543--7548}
  (\bibinfo{year}{2017}).

\bibitem{boers2014prediction}
\bibinfo{author}{Boers, N.} \emph{et~al.}
\newblock \bibinfo{title}{Prediction of extreme floods in the eastern central
  andes based on a complex networks approach}.
\newblock \emph{\bibinfo{journal}{Nature communications}}
  \textbf{\bibinfo{volume}{5}}, \bibinfo{pages}{5199} (\bibinfo{year}{2014}).

\bibitem{boers2015complex}
\bibinfo{author}{Boers, N.}, \bibinfo{author}{Donner, R.~V.},
  \bibinfo{author}{Bookhagen, B.} \& \bibinfo{author}{Kurths, J.}
\newblock \bibinfo{title}{Complex network analysis helps to identify impacts of
  the el ni{\~n}o southern oscillation on moisture divergence in south
  america}.
\newblock \emph{\bibinfo{journal}{Climate Dynamics}}
  \textbf{\bibinfo{volume}{45}}, \bibinfo{pages}{619--632}
  (\bibinfo{year}{2015}).

\bibitem{feng2014north}
\bibinfo{author}{Feng, Q.~Y.} \& \bibinfo{author}{Dijkstra, H.}
\newblock \bibinfo{title}{Are north atlantic multidecadal sst anomalies
  westward propagating?}
\newblock \emph{\bibinfo{journal}{Geophysical Research Letters}}
  \textbf{\bibinfo{volume}{41}}, \bibinfo{pages}{541--546}
  (\bibinfo{year}{2014}).

\bibitem{hersbach2020era5}
\bibinfo{author}{Hersbach, H.} \emph{et~al.}
\newblock \bibinfo{title}{The era5 global reanalysis}.
\newblock \emph{\bibinfo{journal}{Quarterly Journal of the Royal Meteorological
  Society}} \textbf{\bibinfo{volume}{146}}, \bibinfo{pages}{1999--2049}
  (\bibinfo{year}{2020}).

\bibitem{harvey2014equator}
\bibinfo{author}{Harvey, B.}, \bibinfo{author}{Shaffrey, L.} \&
  \bibinfo{author}{Woollings, T.}
\newblock \bibinfo{title}{Equator-to-pole temperature differences and the
  extra-tropical storm track responses of the cmip5 climate models}.
\newblock \emph{\bibinfo{journal}{Climate dynamics}}
  \textbf{\bibinfo{volume}{43}}, \bibinfo{pages}{1171--1182}
  (\bibinfo{year}{2014}).

\bibitem{perkins2017changes}
\bibinfo{author}{Perkins-Kirkpatrick, S.~E.} \& \bibinfo{author}{Gibson, P.~B.}
\newblock \bibinfo{title}{Changes in regional heatwave characteristics as a
  function of increasing global temperature}.
\newblock \emph{\bibinfo{journal}{Scientific Reports}}
  \textbf{\bibinfo{volume}{7}}, \bibinfo{pages}{12256} (\bibinfo{year}{2017}).

\bibitem{rousi2022accelerated}
\bibinfo{author}{Rousi, E.}, \bibinfo{author}{Kornhuber, K.},
  \bibinfo{author}{Beobide-Arsuaga, G.}, \bibinfo{author}{Luo, F.} \&
  \bibinfo{author}{Coumou, D.}
\newblock \bibinfo{title}{Accelerated western european heatwave trends linked
  to more-persistent double jets over eurasia}.
\newblock \emph{\bibinfo{journal}{Nature communications}}
  \textbf{\bibinfo{volume}{13}}, \bibinfo{pages}{3851} (\bibinfo{year}{2022}).

\bibitem{sornette2012dragon}
\bibinfo{author}{Sornette, D.} \& \bibinfo{author}{Ouillon, G.}
\newblock \bibinfo{title}{Dragon-kings: mechanisms, statistical methods and
  empirical evidence}.
\newblock \emph{\bibinfo{journal}{The European Physical Journal Special
  Topics}} \textbf{\bibinfo{volume}{205}}, \bibinfo{pages}{1--26}
  (\bibinfo{year}{2012}).

\bibitem{sachs2012black}
\bibinfo{author}{Sachs, M.}, \bibinfo{author}{Yoder, M.},
  \bibinfo{author}{Turcotte, D.}, \bibinfo{author}{Rundle, J.} \&
  \bibinfo{author}{Malamud, B.}
\newblock \bibinfo{title}{Black swans, power laws, and dragon-kings:
  Earthquakes, volcanic eruptions, landslides, wildfires, floods, and soc
  models}.
\newblock \emph{\bibinfo{journal}{The European Physical Journal Special
  Topics}} \textbf{\bibinfo{volume}{205}}, \bibinfo{pages}{167--182}
  (\bibinfo{year}{2012}).

\bibitem{wang_exploring_2023}
\bibinfo{author}{Wang, S.}, \bibinfo{author}{Meng, J.} \& \bibinfo{author}{Fan,
  J.}
\newblock \bibinfo{title}{Exploring the intensity, distribution and evolution
  of teleconnections using climate network analysis}.
\newblock \emph{\bibinfo{journal}{Chaos: An Interdisciplinary Journal of
  Nonlinear Science}} \textbf{\bibinfo{volume}{33}}, \bibinfo{pages}{103127}
  (\bibinfo{year}{2023}).

\bibitem{zhang2019significant}
\bibinfo{author}{Zhang, Y.}, \bibinfo{author}{Fan, J.}, \bibinfo{author}{Chen,
  X.}, \bibinfo{author}{Ashkenazy, Y.} \& \bibinfo{author}{Havlin, S.}
\newblock \bibinfo{title}{Significant impact of rossby waves on air pollution
  detected by network analysis}.
\newblock \emph{\bibinfo{journal}{Geophysical Research Letters}}
  \textbf{\bibinfo{volume}{46}}, \bibinfo{pages}{12476--12485}
  (\bibinfo{year}{2019}).

\bibitem{cai2012more}
\bibinfo{author}{Cai, W.} \emph{et~al.}
\newblock \bibinfo{title}{More extreme swings of the south pacific convergence
  zone due to greenhouse warming}.
\newblock \emph{\bibinfo{journal}{Nature}} \textbf{\bibinfo{volume}{488}},
  \bibinfo{pages}{365--369} (\bibinfo{year}{2012}).

\bibitem{goyal2021historical}
\bibinfo{author}{Goyal, R.}, \bibinfo{author}{Sen~Gupta, A.},
  \bibinfo{author}{Jucker, M.} \& \bibinfo{author}{England, M.~H.}
\newblock \bibinfo{title}{Historical and projected changes in the southern
  hemisphere surface westerlies}.
\newblock \emph{\bibinfo{journal}{Geophysical Research Letters}}
  \textbf{\bibinfo{volume}{48}}, \bibinfo{pages}{e2020GL090849}
  (\bibinfo{year}{2021}).

\bibitem{meehl2007global}
\bibinfo{author}{Meehl, G.~A.} \emph{et~al.}
\newblock \emph{\bibinfo{title}{Global Climate Projections}},
  chap.~\bibinfo{chapter}{10}, \bibinfo{pages}{747--845}
  (\bibinfo{publisher}{Cambridge University Press},
  \bibinfo{address}{Cambridge, UK}, \bibinfo{year}{2007}).
\newblock
  \urlprefix\url{https://archive.ipcc.ch/publications_and_data/ar4/wg1/en/ch10.html}.

\bibitem{rintoul2018global}
\bibinfo{author}{Rintoul, S.~R.}
\newblock \bibinfo{title}{The global influence of localized dynamics in the
  southern ocean}.
\newblock \emph{\bibinfo{journal}{Nature}} \textbf{\bibinfo{volume}{558}},
  \bibinfo{pages}{209--218} (\bibinfo{year}{2018}).

\bibitem{tsonis2008topology}
\bibinfo{author}{Tsonis, A.~A.} \& \bibinfo{author}{Swanson, K.~L.}
\newblock \bibinfo{title}{Topology and predictability of el nino and la nina
  networks}.
\newblock \emph{\bibinfo{journal}{Physical Review Letters}}
  \textbf{\bibinfo{volume}{100}}, \bibinfo{pages}{228502}
  (\bibinfo{year}{2008}).

\bibitem{donges2009complex}
\bibinfo{author}{Donges, J.~F.}, \bibinfo{author}{Zou, Y.},
  \bibinfo{author}{Marwan, N.} \& \bibinfo{author}{Kurths, J.}
\newblock \bibinfo{title}{Complex networks in climate dynamics: Comparing
  linear and nonlinear network construction methods}.
\newblock \emph{\bibinfo{journal}{The European Physical Journal Special
  Topics}} \textbf{\bibinfo{volume}{174}}, \bibinfo{pages}{157--179}
  (\bibinfo{year}{2009}).

\bibitem{lenton2008tipping}
\bibinfo{author}{Lenton, T.~M.} \emph{et~al.}
\newblock \bibinfo{title}{Tipping elements in the earth's climate system}.
\newblock \emph{\bibinfo{journal}{Proceedings of the national Academy of
  Sciences}} \textbf{\bibinfo{volume}{105}}, \bibinfo{pages}{1786--1793}
  (\bibinfo{year}{2008}).

\bibitem{lenton2019climate}
\bibinfo{author}{Lenton, T.~M.} \emph{et~al.}
\newblock \bibinfo{title}{Climate tipping points—too risky to bet against}.
\newblock \emph{\bibinfo{journal}{Nature}} \textbf{\bibinfo{volume}{575}},
  \bibinfo{pages}{592--595} (\bibinfo{year}{2019}).

\bibitem{nobre2016land}
\bibinfo{author}{Nobre, C.~A.} \emph{et~al.}
\newblock \bibinfo{title}{Land-use and climate change risks in the amazon and
  the need of a novel sustainable development paradigm}.
\newblock \emph{\bibinfo{journal}{Proceedings of the National Academy of
  Sciences}} \textbf{\bibinfo{volume}{113}}, \bibinfo{pages}{10759--10768}
  (\bibinfo{year}{2016}).

\end{thebibliography}
\end{document}